# Architectural Solutions for High-Speed Data Processing Demands of CERN LHC Detectors with FPGA and High-Level Synthesis

Sergei Devadze[1], Christine Elizabeth Nielsen[2], Dmitri Mihhailov[1], Peeter Ellervee[1]
[1]Department of Computer Systems, Tallinn University of Technology, Tallinn, Estonia
[2]National Institute of Chemical Physics and Biophysics, Tallinn, Estonia
{sergei.devadze, dmitri.mihhailov, peeter.ellervee}@taltech.ee, christine.nielsen@kbfi.ee

*Abstract*—The planned high-luminosity upgrade of the Large Hadron Collider (LHC) at CERN will bring much higher data rates that are far above the capabilities of currently installed software-based data processing systems. Therefore, new methods must be used to facilitate on-the-fly extraction of scientifically significant information from the immense flow of data produced by LHC particle detectors. This paper focuses on implementation of a tau lepton triggering algorithm in FPGA. Due to the algorithm's complexity and strict technical requirements, its implementation in FPGA fabric becomes a particularly challenging task. The paper presents a study of algorithm development with the help of High-Level Synthesis (HLS) technique that can generate hardware description from C++ code. Various architectural solutions and optimizations that were tried out during the design architecture exploration process are also discussed in the paper.

## I. Introduction

With the upcoming upgrade to the Large Hadron Collider (LHC) beam [1][2] at CERN, a tremendous growth of the amount of generated measurement data is expected. This will result in extremely high data rates far above bandwidth and technical capabilities of the currently installed data acquisition systems at LHC. In particular, the rate of data coming from particle detectors at the Compact Muon Solenoid (CMS) experiment will be raised from 2Tb/s to 50Tb/s [3]. This massive flow of information about detected particles (events) cannot be processed in real-time by a pure software approach and must be handled with the use of special hardware means.

Most data coming out of the CMS detectors does not represent any interest from an experimental physics point of view. Only a small fraction of detected particles contains relevant information and can be used for further analysis. One particle of particular interest to modern physics research is the tau lepton. However, tau leptons cannot be directly identified by CMS detectors due to their rapid decay into other particles. Instead, they can be found by registering and analyzing their post-decay products [4].

The algorithm that performs tau lepton particle finding and reconstruction of its properties out of information about decay products is known as HPS (Hadron-Plus-Strips) tau lepton triggering [3]. The algorithm consists of multiple steps and involves many mathematical calculations and data manipulation operations. The HPS tau algorithm has previously only been implemented in software and used in post-analyses of reconstructed events. However, it is obvious that real-time data processing with a tau lepton triggering algorithm is not possible in software even if it will run on a top-speed CPU. Instead, a hardware implementation of the algorithm (i.e., via FPGA) is required in order to cope with high rates of incoming data. For this goal, a high-end AMD (Xilinx) VirtexUltraScale+ FPGA has been selected as target hardware for on-the-fly tau lepton reconstruction. This FPGA was chosen by CMS experiment group [2] to be a primary FPGA platform basing on such criteria as high-performance, high-speed I/O count, price/performance ratio, etc.

This paper describes a specific algorithm for the CMS experiment, but the principles and concerns will remain much the same for other tasks related to building FPGA-based designs which handle extremely high data rates. The idea of the paper is to present the approach taken for development of the HPS tau triggering in FPGA hardware. Because of the tight requirements a straightforward implementation of the algorithm was not possible. Instead, various architectural solutions that were tried out during the design exploration process are presented and compared in the paper.

## II. Tau Triggering Algorithm

This section provides background information about the algorithm, its steps, and requirements.

### A. Algorithm Description

The HPS tau lepton triggering algorithm [3] consists of the following major steps (Fig. 1). In the first step all the detected particles are analyzed to select 16 charged particles with the highest transverse momentum (*Pt*) (out of 128 potentially registered particles). The selected particles are called *Seeds*.

Next, for every selected seed, the initial 128 input particles are filtered against the seed and a list of maximum 30 candidates is formed. The filtering is done by selecting particles detected in the vicinity of a particular seed. This implies calculation of distance between the seed and a particle which in its turn requires two multiplication operations and multiple additions/subtractions. In addition, a total transverse momentum (*totalPt*) is calculated (per candidate list).

In the next step, the 16 lists of 30 selected candidates each are filtered again, depending on the relation of every candidate to several criteria like particle type (neutral or charged, hadron, electron, or proton), detected position, location within the area defined by total *Pt*, and other parameters. Every filtering operation demands several multiplications/additions, comparison operations, and checking boolean conditions.

This work was supported by the Estonian Research Council grant PUT PRG780 "Preparing the CMS experiment for high luminosity operations through trigger improvements".

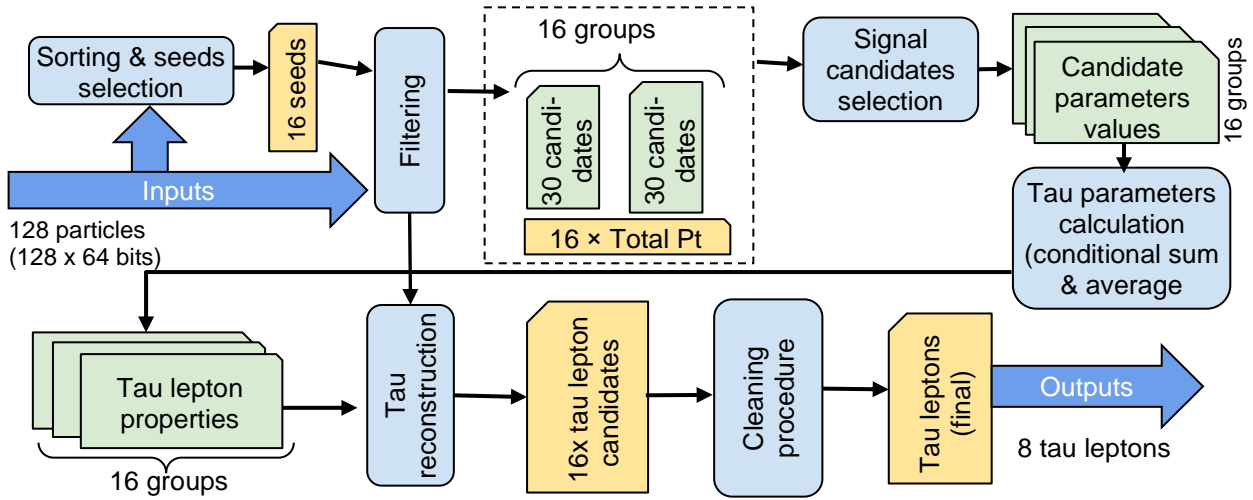

Fig. 1. HPS tau lepton reconstruction algorithm steps

After that, for each of 16 groups, the values of parameters of remaining candidates must be factored with a coefficient, conditionally summed, and then averaged. This involves multiplications, additions and two operations of division (per each group). The resulting values are used as reconstructed properties of an original (decayed) tau lepton.

In the final steps, the algorithm reconstructs 16 tau lepton particles out of calculated properties (some of the reconstructions could be empty, if the algorithm calculations showed that no tau lepton decay occurred). However, as the reconstruction happens out of the same input data, it may occur that two (or more) reconstructed particles are related to the same (original) tau lepton. For that reason, a cleaning procedure is engaged as a last step.

During the cleaning process, sets of particles located close to each other have to be identified. Every pair of 16 particles has to be checked to see if they belong to the same proximity. For every set, a particle with the highest $Pt$ has to be selected while others are dropped (as a repeated/secondary detection of the same tau lepton). Finally, the algorithm produces a list of max 8 truly detected tau leptons.

### B. General Requirements

The general requirements for the algorithm are listed below.

- The algorithm must be implementable on the AMD/Xilinx VirtexUltrascale+ device *xcvu9p* (fit area and timing constraints).
- As the same device is also used for other processing, it is preferable if the algorithm logic fits a single SLR (super-logic region) of this FPGA device (Table I). Spanning across several SLRs is less desirable as it consumes space of other processing modules and requires usage of less-efficient cross-region interconnect resources.
- The overall data processing latency should be no more than 1μS, which is the time window allocated for the tau triggering step of the entire CMS trigger chain. This includes the time needed for data pre-processing, hence the latency of the algorithm itself should be no more than ~0.76μS.
- The new input data comes every 0.15μS, hence by that time the algorithm's pipeline must be capable of starting a new processing iteration. In other words, the initiation interval (II) of the algorithm is 0.15μS.
- The design must communicate with other processing cores (inside the same FPGA) at the frequency of 360MHz (i.e., the rate of readout of the CMS system). This means either the algorithm operates at 360MHz or, if not, implements an extra synchronization.

With respect to the target FPGA the following limitations must be satisfied (Table I and II).

TABLE I. RESOURCE CONSTRAINTS

| Resource type | Available in target FPGA (1 SLR) |
|---|---|
| Logic (LUTs) | 394080 |
| Distributed memory (FFs) | 788160 |
| DSP Slices | 2280 |

TABLE II. PERFORMANCE CONSTRAINTS

| Timing constraints (360MHz clock) | Max allowed clock cycles |
|---|---|
| Latency | 275 |
| Initiation Interval (II) | 54 |

### III. HIGH-LEVEL SYNTHESIS AND RTL DESIGN METHODOLOGIES

Traditionally, FPGA firmware is developed using one of HDL languages (e.g., VHDL or Verilog) and RTL design concept. However, the complexity of the tau triggering algorithm raised questions on the efficiency of this approach.

First, the estimated amount of effort needed to manually create an HDL codebase to perform all needed calculation steps is substantial. Beyond this, the resulting code will be difficult to review and maintain, since the original algorithmic steps defined by their reference implementation of tau triggering in software language have to be transformed into a multitude of FSMs, registers, signal/buses, FIFOs, and other HDL blocks. As there will be no direct mapping between the reference software implementation and RTL code, it complicates introducing any changes into the algorithm when some adaptations or updates are made. This makes implementation of the algorithm using an RTL HDL code impractical.

To overcome the above-mentioned issues, the decision was made to use a High-Level Synthesis (HLS) approach

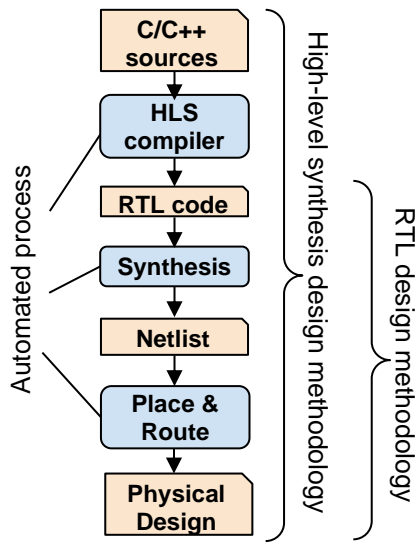

Fig. 2. HLS and RTL design methodologies

[5][6] that is capable of automatically producing RTL HDL out of C++ source code (Fig 2). In this way, one can benefit from usage of typical C++ constructs such as loops, branches, high-level data structures, etc. AMD/Xilinx offers HLS as an integrated part of its Vivado/Vitis Design Suite [7].

Of course, usage of HLS has its own drawbacks. One must follow a certain style of C++ code writing to make this code synthesizable. Moreover, HLS supposes usage of special C++ preprocessor keywords (*#pragma*) to instruct the synthesizer how one or another C++ construct should be translated into HDL code. Examples include unrolling a loop or automatic arrangement of a pipeline. However, it is often difficult to control exactly how the translation is performed, which results in many try-and-see attempts with different code styles and pragma sets before the proper result is finally obtained. The other disadvantage of HLS is its inability to correctly predict the usage of resources in target FPGA and assess delays of the final design. Despite HLS reports providing estimation, they are often far from the actual values calculated after place-and-route. Moreover, for complex designs routing congestion (i.e., inability to perform optimal routing due to concentration of routes within certain areas) could be a major problem. However, potential congestion is not estimated by HLS at all.

Nevertheless, the usage of HLS helps to keep the algorithm sources in clear and compact form, streamlines the design exploration process, ensures ease of algorithm maintenance in future, as well as simplifies experiments with algorithm variations. The latter is especially important as we expect continuous improvements and adaptations of the tau lepton reconstruction procedure.

## IV. CHALLENGES IN IMPLEMENTING ALGORITHM IN FPGA

Implementing the HPS tau algorithm comes with many challenges, both from the technical demands that the LHC beam upgrade necessitates, and from the complexity of the algorithm itself. These include limitations of the FPGA hardware, as well as external constraints such as time window for the design to operate in.

### A. Device-based Design Constraints

The first of these is FPGA performance limits: while FPGA devices have unparalleled flexibility, the inherent trade-off for this is that FPGA devices run on lower speeds (i.e., the logic switching frequency) compared to fixed-architecture CPUs and ASICs. The tau algorithm is expected to reliably operate on the edge of FPGA speed capabilities, making it particularly important that we carefully design each processing step in the algorithm in order to achieve high operating frequency on the underlying hardware [8].

The absolute resource capacity of an individual FPGA device presents another challenge. An FPGA device contains a limited number of computational resources: logic gate equivalents (LUTs), embedded RAM blocks, and digital signal processor engines (DSPs) among others. Moreover, due to the technical constraints of placement and routing, typically no more than 70-75% of available resources can be utilized by the target design. As the target algorithm has a high level of complexity, fitting the design into available resources becomes a challenge even when working with the large FPGA device. As with FPGA performance limits, handling resource capacity is a major challenge.

This also ties into the routing capacity: alongside computational blocks, FPGA designs use interconnection (routing) resources to connect the various on-chip elements together into functioning designs. Due to the data-intensive nature of our algorithm, avoiding bottlenecks due to low interconnection resources is both difficult, and necessary [9]. In particular, we have foreseen potential issues with routing congestion and usage of sub-optimal (detour) paths, both of which may prevent the algorithm from meeting the timing and frequency requirements.

### B. External Limitations and Hardware Implementation

The CMS trigger chain will operate in a very short time window, and each stage within that chain is allocated a fraction of that. One element of handling this is finding a proper balance between the depth of the pipeline and maximal operating frequency of the design. Longer pipelines have the advantage of reducing the combinational logic depth, thus raising possible switching frequencies, but increasing the overall latency. Thus, finding the optimal ratio between the operating frequency and pipeline depth is crucial [10][11].

## V. DESIGN ARCHITECTURE EXPLORATION

The key advantage of using FPGA for high-speed data processing is an unbeatable flexibility of programmable logic and outstanding potential for parallelization. Unlike CPU-based solutions where the ability to perform parallel calculations is limited by a small number of separate CPU cores, FPGAs allow for accommodating many different processing modules all working simultaneously. The latter grants the possibility to arrange massive parallelization and exploit various styles of pipelining. However, to benefit from the advantages of the FPGA platform, a thorough architecture and design space exploration has to be performed to map the required behavior into available programmable logic resources. In this section, we present an iterative approach for design exploration and implementation that was utilized during the development of the HPS tau lepton algorithm.

### A. Automatic RTL Code Synthesis

As the first and straightforward step, the reference C++ implementation of the algorithm in software has been taken "as is" to produce RTL code with the help of Vivado HLS tool-suite. It should be noted that the used C++ code was not initially meant for mapping to hardware and served only for proof-of-concept simulations (in software) and experimental runs on PC workstation. For that reason, the reference algorithm implementation has been adapted in a "rough-and-ready" fashion to fit the requirements of HLS compiler:

- Parameters and datatypes of top (main) function redefined to conform with HLS I/O types and protocols.
- Dynamic memory allocation replaced by statically reserved buffers.
- C++ pointer arithmetic substituted with index-based operations on arrays.
- Vivado HLS was instructed to perform an automatic pipelining of the whole algorithm (i.e., *#pragma hls pipeline* directive used for top-level function) to fit the requirement of initiation interval (II).

This approach had required minimal effort to quickly yield the resulting RTL code. However, the quality of the outcome was far from acceptable:

- The HLS tool was unable to effectively cope with large (non-partitioned) algorithmic C++ description - the synthesis process took more than several days (on a modern high-performance workstation).
- The automatic pipelining produced a design with over **10x** excess in both latency and resource consumption.

The poor results by straightforward usage of HLS synthesis on initial algorithm description were a main driver for further exploration of possible design architectures.

*B. Design Partitioning*

The next step in our implementation approach was to manually partition the algorithm into several separate stages of a pipeline and arrange the communication channels between them. As a result, the following pipeline stages have been identified (Table III). Those stages roughly correspond to the steps of the algorithm described in Fig. 1 (blue boxes).

The HLS tool has been instructed to synthesize every stage separately. FIFO buffers have been used as default channels for passing data from one step of the pipeline to another. The synchronization has been implemented in a dataflow manner, meaning that every stage starts processing independently, as soon as new input data (i.e., from the previous stage) becomes available. The list of stages together with the resources required for their implementation (estimated by Vivado HLS tool) as well as timing parameters are given in Table III. The total reported latency is less than the sum of latencies of all steps since certain pipeline stages run concurrently.

The first step (Stage 1) performs sorting on input data to select the 16 seeds. The implementation of this step is described in detail in [12].

The next two steps are dealing with filtering of the initial 128 input particles (against the seeds), composing lists of maximum 30 candidates (i.e., a separate list for each of 16 seeds), and calculation of a total transverse momentum for every list (*totalPt*). It should be noted that the analysis of all inputs iteratively is not feasible since it requires at least 128 clock cycles hence will violate the initial interval requirement of 54 cycles. Therefore, two separate pipeline stages have been arranged: filtering (Stage 2) and merging (Stage 3). We used a four filtering blocks (for every of 16 seeds) that analyze 32 particles out of 128. In total 4×16 = 64 filtering blocks are required. Stage 3 merges 4x results into a single list (one list for every seed), as described in Section V(C).

TABLE III. DESIGN PARTITIONING RESULTS

| # | Pipeline stages (steps) | Resource usage | | | Timing | |
|---|---|---|---|---|---|---|
| | | LUT | FF | DSP | Latency | II |
| 1 | Seeding | 10250 | 4874 | 0 | 43 | 43 |
| 2 | Filtering (4x blocks) | 11103 | 12880 | 64 | 38 | 38 |
| 3 | Merging 4x32→1x30 | 18276 | 6345 | 0 | 38 | 34 |
| 4 | Signal candidates selection | 14967 | 11474 | 16 | 37 | 36 |
| 5 | Tau parameters calculation | 15719 | 13578 | 32 | 59 | 35 |
| 6 | Tau reconstruction | 3195 | 706 | 0 | 1 | 1 |
| 7 | Tau cleaning | 25575 | 18242 | 184 | 13 | 13 |
| | Pipeline arrangement, inter-stage FIFOs, control, etc. | 41794 | 30141 | 0 | 8 | |
| | **Grand total** | 140879 | 91895 | 296 | 203 | 44 |
| | % of SLR | 35 | 11 | 4 | | |

Stage 4 analyzes all 16 lists to identify signal candidates. This involves calculating the proximity of every candidate with respect to the *totalPt* values from Stage 2 and other conditions. Stage 5 performs a conditional sum and averaging of the resulting values which are then used to reconstruct tau leptons (Stage 6). Stage 5 consumes a significant amount of resources due to the division operations used for averaging. The last step (Stage 7) is discussed in Section V(D).

Despite HLS synthesis of the partitioned design yielded much better results than the attempt described in Section V(A), the placement and routing was still failing to satisfy the signal timing requirements (see column (1) in Table VI). The reported total negative slack (TNS) was *-43604ns* and the worst negative slack (WNS) was *-1.297ns*. The detailed analysis of timing report revealed that most of the problems are related to the logic blocks of steps 1, 2, 3 and 7.

To improve the synthesis results, we have examined pipeline stages for potential improvements. In this paper, optimization of steps 3 and 7 is discussed. The optimizations applied for step 1 are presented in [12]. The optimization of step 2 is planned for future.

*C. Optimization of the Merging Step*

The goal of the merging procedure is to compose a single list of candidates out of four separate lists that have been obtained during the previous step of filtering. Each of the source lists may contain up to 32 items (the exact count varies since the filtering could drop out some/all of them). The target list storage is implemented as a FIFO buffer. The merging must place a maximum of 30 candidates into this buffer. If the total number of items exceeds this capacity, there is no preference which items are chosen as target candidates.

*a) Merging Solution A (original)*

The storage for the source lists is implemented as four FIFO buffers with a depth of 32. The number of items in every buffer is known beforehand and is stored in four *Size* registers ($S_i$, *i=1..4*). Those registers are required to distinguish the items that belong to the current iteration of the algorithm (not to mix with the items added as a part of the next iteration).

Consequently, the straightforward approach would be to read out every FIFO until either the target FIFO is filled or there are no more source items left. However, we cannot simply stop the processing after reaching 30 target candidates, since then there will be leftovers in the source FIFOs that will ruin the overall algorithm's pipeline and synchronization. Neither can we reset the FIFO completely, as it may already contain the items of the next iteration cycle (that must be preserved). Therefore, all items that belong to the current iteration of the algorithm (including excessive ones) should be explicitly read-out (removed) from the source FIFOs. A simple iterative removal of leftovers is not possible as it takes up to 128 cycles (hence violates the initiation interval policy).

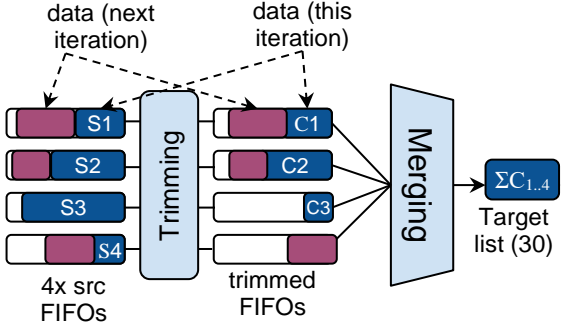

Fig. 3. Merging solution A (FIFO buffers and trimming)

To overcome the issue, the merging was divided further into two sub-steps (Fig. 3):

1. Trimming. This sub-step is needed to explicitly remove the excess of items from the source FIFOs. First, we calculate how many items ($C_i$, $i=1..4$) should be taken from every source FIFOs to avoid overflow of the target list. Next, the items from every of the source FIFOs are then read $C_i$ times and are placed into intermediate FIFOs. The remaining ($S_i$-$C_i$) items are simply read and dropped out. The reading is implemented in parallel for all source FIFOs thus never exceeds 32 cycles.
2. Then merging is performed straightforwardly. Each of the intermediate FIFOs is consequently read as many times as the corresponding value of the $C_i$ register. The read item is placed into the target FIFO.

Both sub-steps are run in a pipeline, i.e., the second step runs after the first items are put into the intermediate buffer (hence preserving the overall initiation interval policy).

   *b) Merging Solution B (optimized)*

In this version, we have implemented source storage lists as four separate Ping-Pong Buffers (PIPO) instead of FIFO. Similar to *Solution A*, four *Size* registers ($S_i$, $i=1..4$) were used to keep the number of items in every PIPO buffer. Items in the source PIPOs are stored adjacent to each other, the buffer may contain unused (empty) space.

Since PIPO buffer is capable of random-access (unlike sequentially accessed FIFO), it has no problem with leftovers (the unused data will be overwritten on the next algorithm's iteration). However, the merging step should be capable to process the items before they will be overwritten by the next algorithm's iteration. To arrange concurrent reading of all PIPOs and selection of candidates, the following procedure has been proposed (Fig. 4).

An *Index* register is used to store addresses of currently processed items in the source PIPO buffers. A single instance of *Index* register serves for all 4 source PIPO buffers. Initially *Index* is set to 0. The *Count* register tracks the number of already processed target candidates (set to 0 at start). Four 1-bit *Available* registers ($A_i$ $i=1..4$) are implemented to mark the availability of an item in each of the source PIPOs.

The merging is executed as follows (Fig. 4):

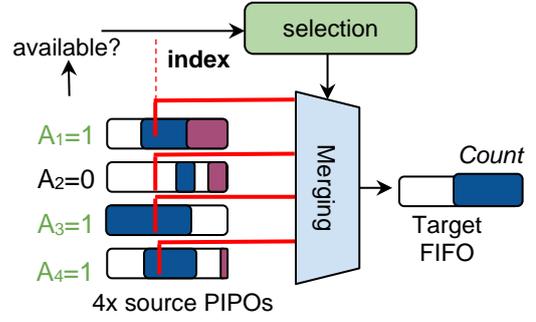

Fig. 4. Merging solution B (with PIPO buffers)

1. The value of the *Index* register is compared with each of the *Size* registers ($S_i$). If the *Index* is greater than the number of elements in source list ($S_i$) - the corresponding $A_i$ register is set to *0* (otherwise *1*).
2. All four $A_i$ registers are evaluated. If none of them contains *1* value - the merging is done (no more source items). Otherwise, the item that corresponds to first $A_i$==1 register is taken from the source list and placed to target FIFO. In the latter case, the corresponding $A_i$ is set to *0* and *Count* is increased.
3. If *Count* is 30, then merging is done (target list is full).
4. When all $A_i$ registers are equal to *0*, then the *Index* register is increased, and execution continues from step 1. Otherwise, processing continues from step 2.

We have compared both solutions (Table IV) in terms of resource usage and latency properties. *Solution B* has been chosen as it consumes considerably less logic resources and offers better performance.

TABLE IV. MERGING STEP OPTIMIZATION RESULTS

| Merging step | Solution A | Solution B |
|---|---|---|
| Latency, clock cycles | 38 | 33 |
| Initiation interval, clock cycles | 34 | 33 |
| Logic (LUTs) | 18276 | 13554 |
| Distributed memory (FFs) | 6345 | 6376 |

*D. Optimization of the Tau Cleaning Step*

The goal of the cleaning step is to inspect the list of 16 final candidates and get rid of repeated (secondary) detections of the same tau leptons. For any group of tau candidates that have similar coordinates, the algorithm has to select the tau which has the highest transverse momentum *Pt* (as a truly detected tau lepton). The other candidates are considered as secondary detections of the same tau lepton and must be dropped. At maximum 8 tau leptons can be selected.

   *a) Tau Cleaning Solution A (original)*

The initial implementation performs cleaning in two steps:

1. Candidates are graded descending according to their *Pt*. The candidate grading is implemented in hardware via a simple sorting network.
2. For every *i*-th candidate (*i=1..16*), we compare its coordinates with the *j*-th candidate, where *j* is from *i+1* to *16*. If the coordinates of a candidate pair *(i, j)* belong to the same proximity, the *j*-th candidate is marked as dropped. As the candidates were sorted beforehand, the selection of the one with the highest *Pt* is guaranteed.

   *b) Tau Cleaning Solution B (optimized)*

In this implementation we do not perform explicit candidate sorting. Instead, we use the following sub-steps:

1. Build a cleaning matrix *M* with size of 16×16. For every candidate pair *(i, j)* we find out two properties:

   *NearBy* is *1* iff the coordinates of those two candidates are close to each other.
   *LessPt* is *1* iff candidate *i* has less *Pt* value than candidate *j*.

   The resulting matrix is constructed as follows:
   M(i,j) = NearBy(i,j) AND LessPt(i,j)

2. Determine candidates that must be dropped using the expression below. For any *i*-th candidate:

   ∀ M(i,j)=1 (where i<>j) ⇒ Drop out i-th candidate

The illustration of the method is given in Fig 5. The table on the left lists an example of 6 taus, their proximity groups and *Pt* values. If two items belong to the same proximity, they are located close enough to each other (*NearBy = 1*). The table on the right presents a cleaning matrix for the example. The last column shows taus that have been cleaned out.

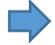

Fig. 5. Example of tau cleaning solution B

*c) Comparison*

Both solutions were compared in terms of resource usage and performance (see Table V). *Solution B* has finally been chosen since it consumes less resources and holds the same initiation interval. Despite *Solution B* has a bit longer latency (+2 cycles), this was considered negligible.

TABLE V. TAU CLEANING STEP OPTIMIZATION RESULTS

| Tau cleaning step | Solution A | Solution B |
|---|---|---|
| Latency, clock cycles | 13 | 15 |
| Initiation interval, clock cycles | 13 | 13 |
| Logic (LUTs) | 25575 | 22467 |
| Distributed memory (FFs) | 18242 | 9685 |
| DSPs hard-macros | 184 | 58 |

*E. Exploring the frequency and latency trade-off*

As a part of the architecture exploration, the design has been examined to work on different operating frequencies. As a matter of fact, working on higher frequency implies difficulties in meeting signal timing requirements, while lower operating frequency restricts the amount of clock cycles available for completion of data processing.

The nominal target frequency was set to 360MHz due to the external constraints (corresponds to the frequency of other processing modules inside the same FPGA). However, after placement and routing, it became evident that the design was not meeting signal timing requirements to operate at such speed (see TNS/WNS values of column (2) in Table VI).

One of the viable solutions to overcome this issue is to try to lower the operating frequency. The latter is possible since the latency and initiation interval of the implemented algorithm were less than allowed thresholds. At the same time, operation at another frequency required extra synchronization logic between the two clock domains which added overhead both in latency and resource usage.

The frequency of 300MHz has been selected because it is the lowest possible speed which still fits the initiation interval (II), i.e., pace of incoming data. The required II is 0.15uS (54 cycles at 360MHz) corresponds to 45 cycles if run the logic at 300MHz (achieved II of the algorithm). Despite the implementation required extra 10 cycles for clock domains synchronization, it still fits the latency requirements.

The results of the algorithm implementation with reduced operational speed were satisfying all the requirements including timing constraints (see column (3) in Table VI). The layout and the routing congestion metrics for the non-optimized (running at full 360MHz speed) and the optimized (300MHz) implementations of the algorithm are shown in Fig. 6 (dark orange color represents the areas with the most congested routing, blue areas mean no congestion).

TABLE VI. IMPLEMENTATION RESULTS

|  | (1) | (2) | (3) |
|---|---|---|---|
| *Implementation version* | *Initial* | *Optimized* | |
| *Operating frequency, MHz* | *360* | *360* | *300* |
| Max allowed latency, clock cycles | 275 | 275 | 220 |
| Target initiation interval, clock cycles | 54 | 54 | 45 |
| Achieved latency, clock cycles | 203 | 200 | 210 |
| Achieved initiation interval, clock cycles | 45 | 45 | 45 |
| Worst negative slack (WNS), ns | -1.297 | -0.576 | 0.02 |
| Total negative slack (TNS), ns | -43604 | -2451 | 0 |

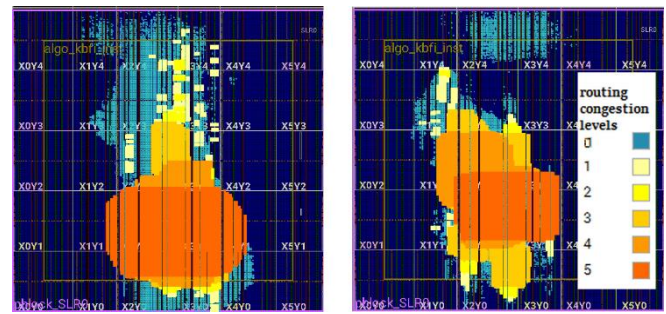

Fig. 6. Layout and routing congestion: initial 360MHz (left) and optimized 300MHz (right) design versions

## VI. CONCLUSION

In this paper, we presented a case study of FPGA-based design of the HPS tau triggering algorithm to be used at the CMS experiment at CERN. The hardware implementation of the algorithm was developed with the help of the High-Level Synthesis methodology, which allows rapid translation of high-level programming language description (i.e., C++) into RTL Verilog or VHDL code. Due to strict area, latency and timing constraints coming from the requirements of the CMS, the implementation of the algorithm in FPGA fabric required thorough exploration of various design architectures. The paper presented steps we have taken in our development process, starting from the straightforward mapping from C++ to RTL code, followed by partitioning of the algorithm into pipeline and optimizing the pipeline stages. Despite the paper describes an implementation of a specific HPS tau triggering algorithm, the principles and concerns remain much the same for other FPGA-based designs which required to handle extremely high data rates.